\documentclass[aip,rsi,reprint,amsmath,amssymb]{revtex4-1}

\usepackage[latin1]{inputenc} 
\usepackage[T1]{fontenc}      
\usepackage[pdftex]{graphicx}
\usepackage{color}

\begin{document}


\title{A setup for fast cooling of liquids in sealed containers and at intermediate temperatures} 



\author{D. Schulze Grachtrup}
\email[]{d.s.g@gmx.de}
\homepage[]{www.ipkm.tu-bs.de}
\affiliation{Institut für Physik der Kondensierten Materie, TU Braunschweig, Mendelssohnstraße 3, 38106 Braunschweig, Germany}
\author{M. Kraken}
\affiliation{Institut für Physik der Kondensierten Materie, TU Braunschweig, Mendelssohnstraße 3, 38106 Braunschweig, Germany}
\author{N. van Elten}
\affiliation{Institut für Physik der Kondensierten Materie, TU Braunschweig, Mendelssohnstraße 3, 38106 Braunschweig, Germany}
\author{S. Süllow}
\affiliation{Institut für Physik der Kondensierten Materie, TU Braunschweig, Mendelssohnstraße 3, 38106 Braunschweig, Germany}

\date{\today}

\begin{abstract}
We present a simple layout of a fast cooling system for liquids in sealed containers utilizing the large temperature gradients of cold nitrogen gas. Our system is optimized for about 20 cylindrical containers of 500 cm$^3$ but the setup allows for simple up- and downscaling as well as the use of arbitrary containers. We have characterized the cooling performance of our system experimentally for liquid temperatures in the range from room temperature down to the freezing point at $\approx$ -2 $^{\circ}$C. With our system we achieve container cooling times as low as 3 min, a significant reduction if compared to cooling times with common methods in the range of 8 to 40 min. Modeling of the cooling process proves that convection within the liquid is crucial for quick heat transfer and that the most important factor limiting the cooling rate is the thermal conductivity of the container material.
\end{abstract}


\maketitle 

\section{Motivation}

Generations of low temperatures physicists have cooled their beer using liquid nitrogen\cite{theunissen94,amitsuka13}. The details of the common cooling procedures vary from Lab to Lab, but generally these techniques utilize liquid nitrogen together with large water buckets to produce water-ice mixtures which then serve as a temperature bath at 0\,°C for cooling of the samples which are, for instance, beer containing glass bottles or PET bottles from soft drinks. This way, temperature gradients between sample and the surrounding bath are kept low, to the effect that strain within the container material is minimized, and which in case of direct contact to liquid nitrogen would cause material failures. However, the cooling rate of the container liquids is dominated by the temperature gradients to the bath and thus limited due to the thermodynamic properties of the water-ice mixture. In this configuration, there is room for massive improvement of the cooling process.

In contrast to the traditional approach, we have developed a method that utilizes only cold nitrogen gas for cooling. This way, we combine the advantages of very high temperature gradients, which cause a strongly increased heat flow from the sample to the temperature bath in the cooling process, together with a low heat capacity of the coolant, keeping strain in the container material well below the breaking limit.

In the following we will present a detailed characterization of our cooling setup. We have determined experimentally the temperature profiles of the system and within the samples during the cooling process, and combine these with a numerical simulation of the corresponding time dependent temperature distributions. Comparison with other common cooling techniques shows that our technique produces a reduction of cooling time of the container liquids from room temperature to a drinkable $\approx$~8~°C from 8~-~40~min down to 3~min.

A similar procedure, Flash freezing, is well known and widely applied, especially in the food processing industry\cite{chapman30, sun01}. In this process the sample is also exposed to very low temperatures often created by cold air, dry ice or evaporated liquid nitrogen. However, the big difference to our setup is that in the process of Flash freezing the liquid to solid phase transition is intended and large temperature gradients within the sample do not cause complications. This technique thus requires similar resources but no or hardly any control of the details of the cooling process.

In contrast, with our setup we attain sufficient control of the temperature gradients and distribution to avoid phase transitions and achieve uniform cooling. This way, samples with particular sensitivity to phase transitions, as for instance drinks or vaccines, can be cooled quickly close to their freezing point without taking damage. Based on our setup future developments seem possible utilizing the achieved level of process control. For instance, flow or counter-flow cooling systems without the need for electrical power could be easily developed, and which could satisfy the needs of specialised cooling applications.

\section{Construction}

Heat flow according to Fouriers law is dominated by the temperature gradients\cite{kittel}, i.e.,
\begin{equation}
\vec{q}= k \, \nabla T
\label{eq:fourier}
\end{equation}
for the heat flow density $\vec{q}$ with the thermal conductivity of a material, $k$, and temperature gradient $\nabla T$. Therefore, in order to maximize the temperature gradient $\nabla T$ in Eq. \ref{eq:fourier} and thus the heat flow density $\vec{q}$ one may use nitrogen as a coolant, as it is affordable and easily available in science laboratories using low temperatures. Colder coolants such as helium or hydrogen would provide even higher temperature gradients, but only at much higher cost and increased handling hazards. Therefore, nitrogen is the ideal coolant for fast cooling of consumable liquids in low temperature laboratory environments.

First experimental attempts in fast cooling revealed that common glass as the standard container material of our samples is too brittle and sensitive to allow for very fast temperature changes. In effect, this leads to instant failure of glass containers due to the mechanical stress induced by the thermal gradients when direct contact between liquid nitrogen and glass container is allowed during cooling. In this situation, a simple solution to provide large temperature gradients while keeping the maximum temperature gradients in the container material below the breaking limit is to use cold nitrogen gas at 77 K as coolant instead of liquid nitrogen. Since the gas has a much lower heat capacity and is thus quickly heated by the warm container surface temperature gradients, mechanical strains are sufficiently reduced to prevent breaking of the containers. In exchange for this aggregate induced cooling power limitation we need to provide a steady gas flow to keep the temperature gradients large for efficient cooling.

\begin{figure}
\raisebox{2cm}{\includegraphics[width=0.225\textwidth]{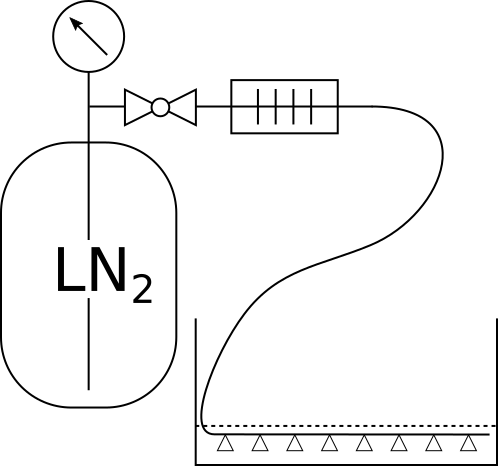}}
\hspace{0.02\textwidth}
\includegraphics[width=0.225\textwidth]{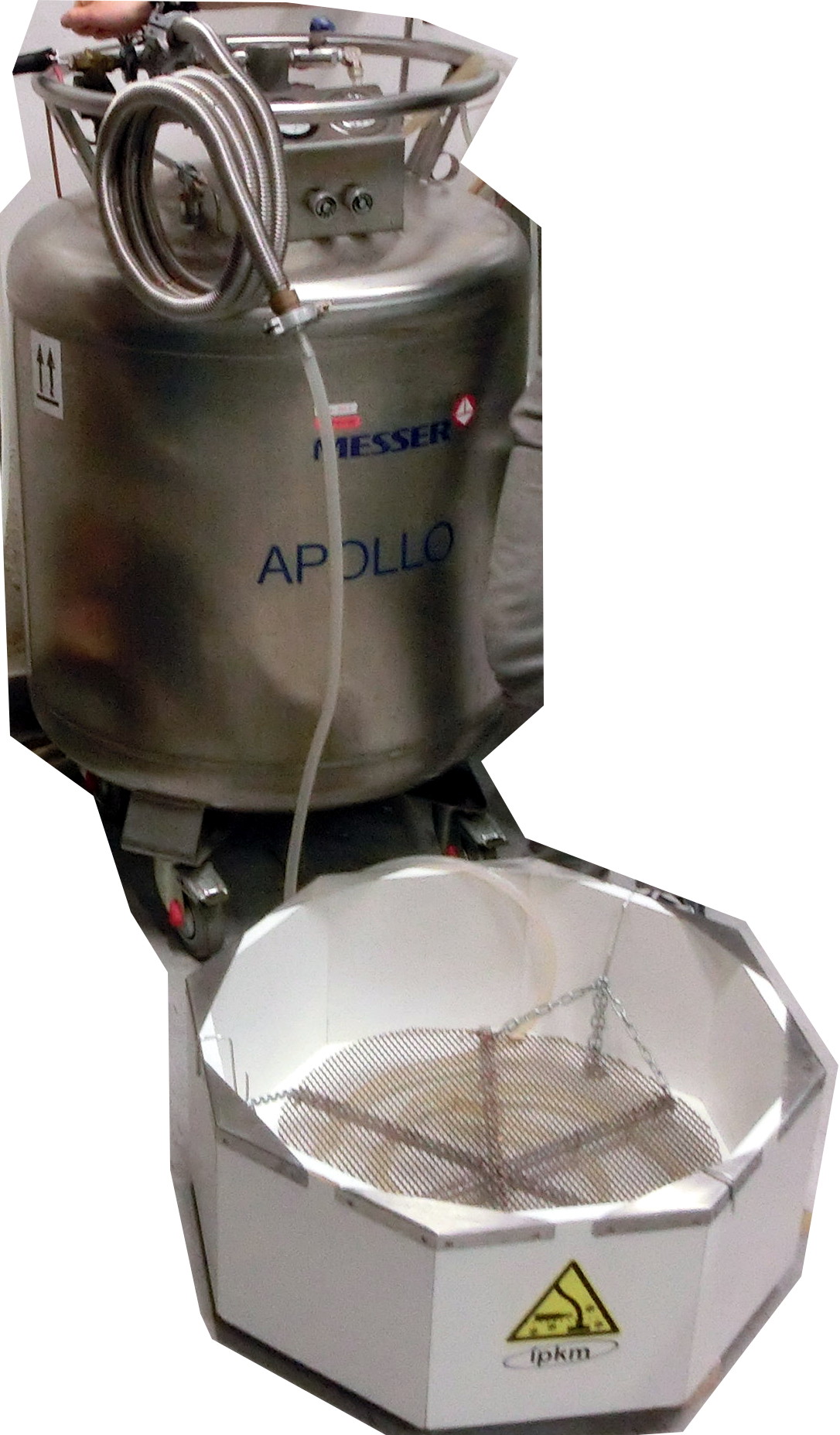}
\caption{(color online) Schematic drawing (left) and experimentally realized (right) setup for the fast cooling of container liquids: The liquid nitrogen (LN$_2$) is stored in a pressurized vessel with attached manometer at 1 bar. The flow is regulated by a ball valve, regulating the gas into a thin-walled flexible steel tube serving as evaporator. The cold nitrogen gas is then transferred via a silicone hose into the cooling setup where it is released in downward direction (away from the containers) by small nozzles (indicated as triangles) mounted under a metal grid (dashed line). The metal grid also serves as sample support. \label{fig:setup-complete}}
\end{figure}

Our setup for fast cooling of liquids in containers is depicted in Fig. \ref{fig:setup-complete}. The main components are a 200 l liquid nitrogen dewar as reservoir to store the nitrogen at a pressure of about 1 bar. The liquid nitrogen is then transferred through a ball valve to a thin-walled flexible steel tube which serves as an evaporator and provides the cold nitrogen gas needed for cooling. From the evaporator the cold nitrogen gas is transferred through a silicone hose into the cooling chamber.

The octagonal cooling chamber with an inner diameter of 60~cm and a height of 24~cm is constructed from chip board. This material is very tolerant to  regular fast thermal cycling and has a low thermal conductivity. This way minimum temperatures close to 77 K can be reached on the inside while touching the outside of the cold cooler is nonhazardous.

Inside the cooling chamber a metal grid is installed at approximately 4~cm above the ground which serves (i) as a sample support and (ii) as a mounting platform for the silicone hose delivering the cold gas. The silicone hose is wound as a spiral below the metal grid and small holes are cut in the hose in downward direction, which serve as nozzles. This way the cold gas is released away from the sample and remaining droplets of liquid nitrogen escaping the evaporator do not touch the sample containers. As well, the cooling is realized only by the use of cold nitrogen gas since such droplets remain at the bottom of the cooling chamber.

The gas flow rate is regulated by cautious adjustment of the ball valve so that a steady gas flow is established with as little (ideally: none) liquid nitrogen escaping the nozzles as possible. After 2 - 3 minute precooling of the setup with increased gas flow the cooling setup has reached equilibrium temperature and the gas flow rate can be strongly reduced and maintained on a low level as long as necessary, providing the steady flow of cold nitrogen gas close to 77 K.

The main parts of our setup, the evaporator and the cooling chamber, can easily be scaled up or down to match the need for bigger or smaller cooling spaces.

\section{Experimental Setup}

We have characterized the above described fast cooling setup by taking spatially resolved temperature profiles of the empty and filled cooling chamber as well as within a filled sample container. All temperature measurements were done using 5 type T (Cu-CuNi) thermocouples with one soldering point immersed in a water-ice mixture. The thermoelectric voltages were measured simultaneously in 2 wire configuration using several Keithley multimeters. All measured voltages were converted to temperatures using a 9th order polynomial fitted to the ITS-90 temperature tables which results in a neglegible numerical conversion accuracy of $\Delta T < 20 \textrm{mK}$.

Residual temperature variations were treated differently for measurements of the cooling chamber (i) and within the liquid (ii). In cooling chamber measurements (i) we have measured the well-defined temperatures of the water-ice mixture used as thermocouple reference at 0~°C and of liquid nitrogen at 77~K. For the liquid measurements (ii) the container wall temperature was measured with an external IR thermometer in thermal equilibrium. In both cases thermocouple voltage readings were shifted appropriately to reproduce these well-defined temperatures. Hence, our overall temperature measurement accuracy is well below 1~K.

\subsection{Cooling chamber}

\begin{figure}
\includegraphics[trim=1cm 1cm 2cm 2cm, width=0.48\textwidth]{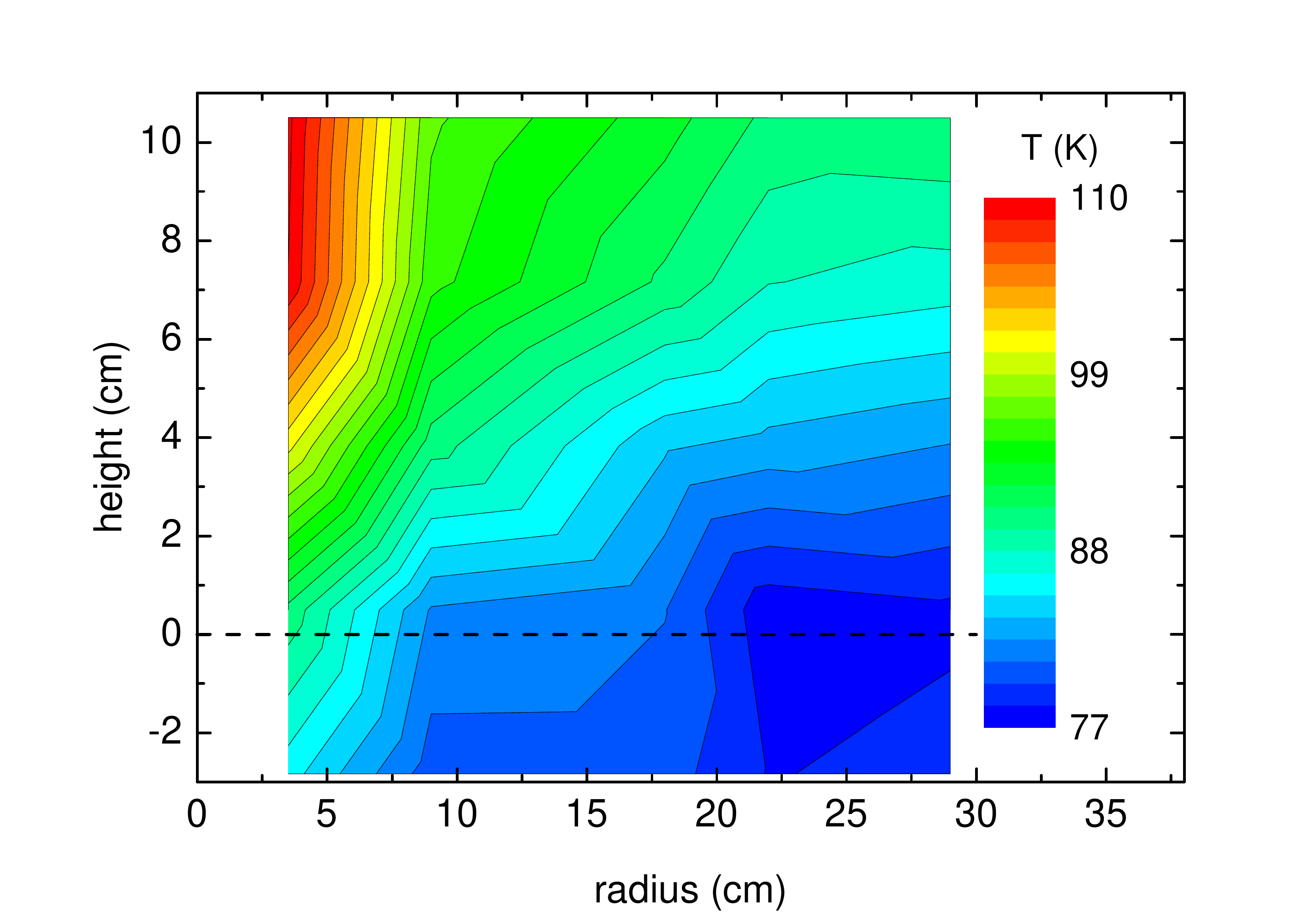}
\caption{(color online) Experimentally determined radial temperature profile inside the cooling setup before inserting sample containers. The dashed black line indicates the position of the metal grid atop the silicone hose with the downward nozzles and supporting the samples. The high temperatures for large height $h$ / small radius $r$ together with low temperatures at small $h$ / large $r$ indicate a vortex formation of the flowing coolant as it is observed by eye; for details see text.\label{fig:TvsPos-cooler}}
\end{figure}

In various measurements of the cooling chamber temperature distribution a typical temperature profile was reproduced reliably (see Fig. \ref{fig:TvsPos-cooler}). For the measurements, we have determined the local temperature in the cooling chamber as function of the radial distance $r$ to the center of the cooler, and the height $h$ above the bottom. Maximum deviations from this profile are found for small radii ($r < 5 \textrm{ cm}$) with $\Delta T \approx $ 5 K for $h \leq 6 \textrm{ cm}$ and rising up to $\Delta T \approx$ 8 K for $h > 6 \textrm{ cm}$. For a larger radius ($r \geq 5 \textrm{ cm}$), which covers about 97\% of the cooler volume, we find all temperature variations from the average values at identical positions to be less than 4 K. We attribute these residual temperature fluctuations to the slight asymmetry of the nozzle positions and slightly different nitrogen gas flow rates which lead to minimal changes in the vortex formation as discussed below. Thus, we have a very stable and reproducable temperature distribution in our setup, and which represents the thermal bath in our simulations (see below). 

During nitrogen gas flow in the cooler, a vortex formation can be observed by eye during precooling or in the temperature profile shown in Fig. \ref{fig:TvsPos-cooler}. It is most likely a result of the spiral-like wound silicone hose. In our experiments, we have found that this vortex formation is very desirable as it stabilizes a constant gas flow of the coldest nitrogen gas at $\approx$~77 K for small $h$ and large $r$ which covers most of the cooler volume.

Altogether, judging from the temperature distribution map, further significant optimization of the cooling procedure by increasing the temperature gradient or changing the temperature distribution seems hardly possible while using cold nitrogen gas as coolant. Since we find a large radial range and only a small range of height with lowest temperatures the most efficient cooling alignment of the sample containers is at low $h$ / large $r$, thus the liquid containers lying with the larger bottom parts pointing outside. This arrangement was used in all tests and measurements presented below.

\subsection{Sample container}

\begin{figure}
\includegraphics[width=0.16\textwidth]{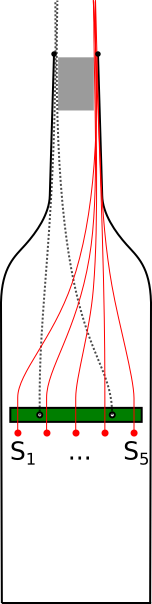}
\hspace{0.02\textwidth}
\includegraphics[width=0.29\textwidth]{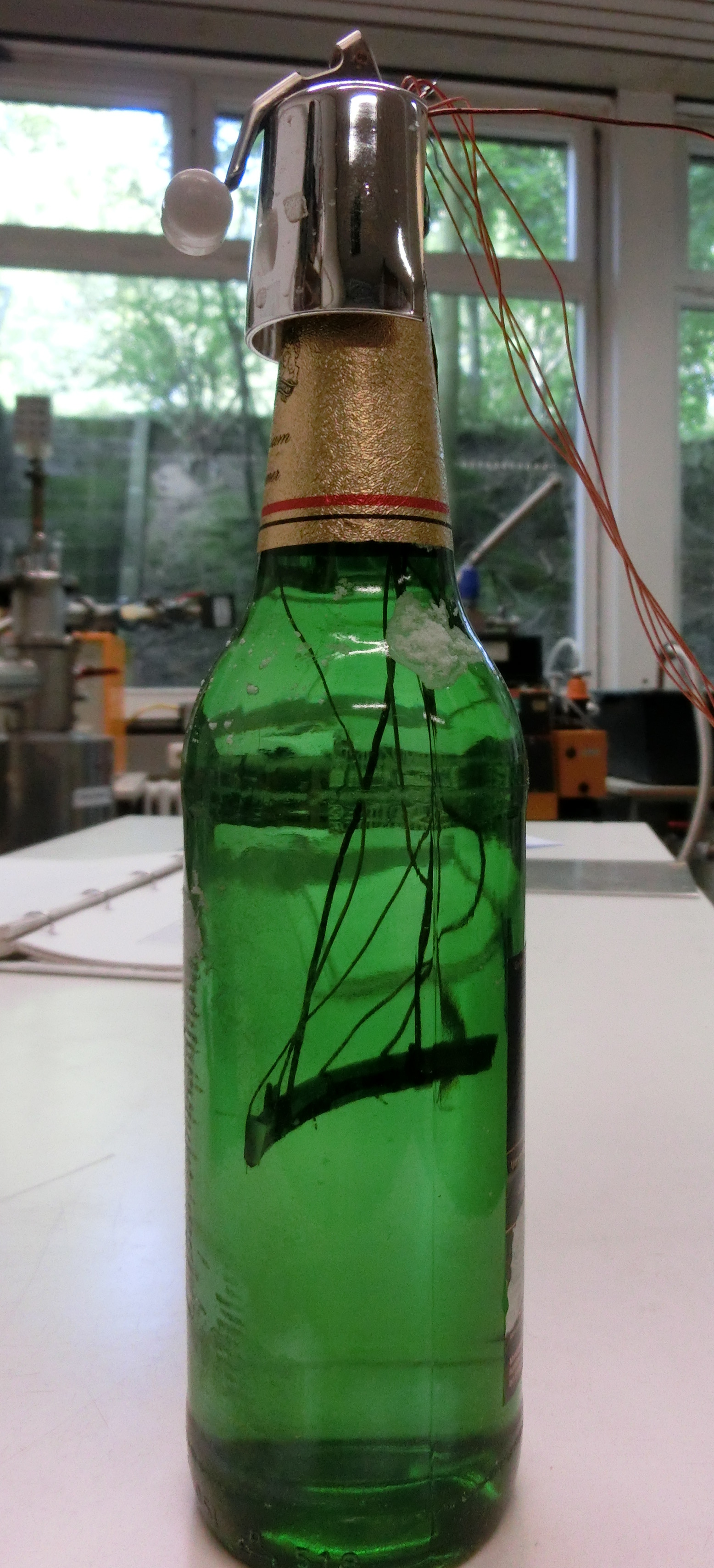}
\caption{(color online) Schematic drawing of the Wolters standard equipped with 5 temperature sensors (left) and the corresponding experimental realization (right). The five thermocouples ($S_1 \ldots S_5$) are mounted on a plastic support (green) with the temperature sensitive tips (red dots) floating freely within the liquid. The support is fixed in position by 2 rigid wires (dark gray, dotted) while the thin sensor wires (red) are mounted without strain. All wires are held firmly in position by the rubber seal (gray) closing the top of the bottle. The liquid consists of 95\% water and 5\% ethanol.\label{fig:setup-beer}} 
\end{figure}

For a quantitative analysis of the efficiency of our cooling setup we have equipped a glass bottle (which we will refer to as the ``Wolters standard'' (Ref. \footnote{For all measurements presented here and most tests during the development of the cooler we have used a 0,5 l bottle Pilsner beer from the local brewery ``Wolters''.})) with the 5 thermocouples which are labeled sensors $S_1$ to $S_5$ subsequently (see Fig. \ref{fig:setup-beer}). The sensors were fixed onto a small plastic support which was inserted in the bottle using two rigid supporting wires. The sensors were positioned in the middle of the widest part of the sample container (the standing bottle) with different radii measured from the middle as depicted in Fig. \ref{fig:setup-beer}. This way, sensors $S_1$ and $S_5$ are closest to the outside but at almost identical radii, sensor $S_3$ is positioned at the center of the container and sensors $S_2$ and $S_4$ in between. After installation of the sensors the bottle was filled with a mixture of 95\% of water and 5\% ethanol and sealed with a rubber seal. A sample of this composition should very closely resemble beer in its thermal properties since water and ethanol are the main components and residual ingredients are only important for the taste. However, real beer is pressurized in the bottle while our sample is not. We expect this difference to be negligible since the saturated vapour pressure of most beers is below 3 bar\cite{beer-pressure} and for the main component, water, the dependence of the freezing point upon pressure is very small. Thus, the reduction of the freezing point down to $\approx$ -2 °C due to the added ethanol is the most dominant modification to the thermal properties of the container liquid.

We have measured two types of temperature profiles in the bottle, i.e., in ``horizontal'' and ``vertical'' alignment. For the horizontal alignment the Wolters standard is positioned in the cooling chamber as described above, that is with the bottom pointing outside. In addition, all sensors are at the same height above the supporting metal grid. Hence, the plastic support holding the thermocouples is parallel to the grid. For the vertical alignment the Wolters standard is rotated by 90° so that sensor $S_1$ is at the bottom, closest to the supporting grid. Thus, we expect sensors $S_1$/$S_5$ and $S_2$/$S_4$ to show identical readings in horizontal arrangement and a successive increase in vertical arrangement during cooling within the cooler.

\begin{figure}
\includegraphics[width=0.48\textwidth]{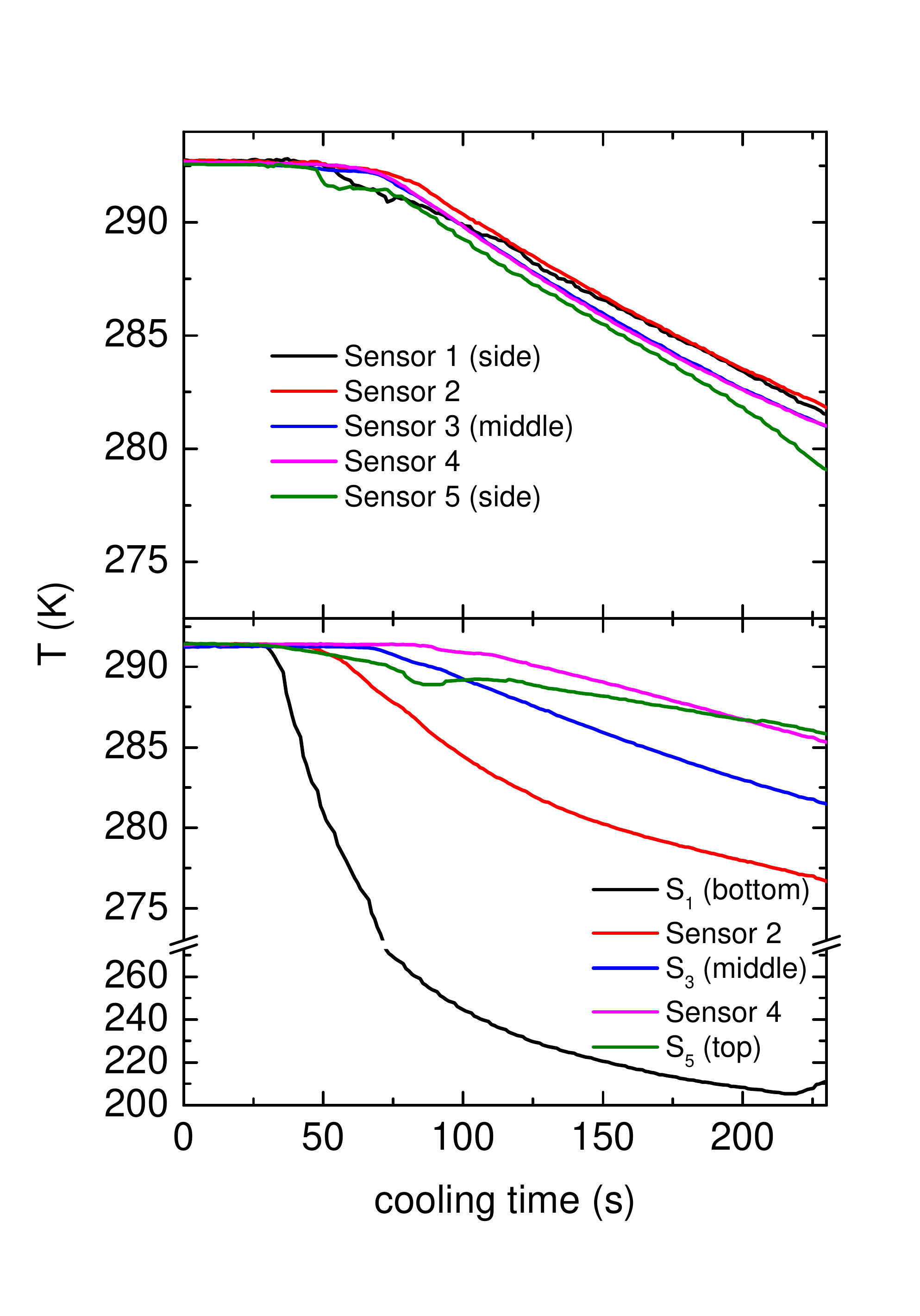}
\caption{(color online) Temperature evolution during fast cooling measured via 5 thermocouples as a function of cooling time for the temperature sensor equipped Wolters standard. The sensors are aligned horizontally (top) and vertically (bottom), respectively; for details see text.\label{fig:Tvst-5sensors}}
\end{figure}

For a typical cooling procedure as described in the preceeding section we find the temperature curves as depicted in Fig. \ref{fig:Tvst-5sensors} for the horizontal and vertical arrangement of the temperature sensors, respectively.

For both arrangements we observe no or only very small temperature changes for the first 50~s, which reflects the low thermal conductivity of the glass container limiting the reaction time of the system. After 50~s the container has been cooled and heat is transferred from the container to the bath leading to a reduction of the container temperature.

For the horizontal alignment (Fig. \ref{fig:Tvst-5sensors}, top) we find very similar cooling curves for all five sensors. This is as expected for the sensor couples $S_1$/$S_5$ and $S_2$/$S_4$, but surprisingly also between both couples of sensors and sensor $S_3$. The similarity of the temperature evolution for all five sensors indicates that the heat transferring mechanism can not be thermal conduction within the liquid alone since in such a scenario the outer sensors would have to cool down prior to the inner ones. Instead, the heat flow within the liquid needs to be strongly increased. Such an increase might be realized by convection within the liquid, which seems reasonable due to the large temperature gradients involved leading to convection. The measurement was stopped after $\approx$ 260 s when significant ice formation was observable on the container walls to prevent breaking of the Wolters standard. We attribute the slightly increased cooling of sensor $S_5$ for $t$ > 200 s to asymmetric ice formation on the container walls.

For the vertical alignment (Fig. \ref{fig:Tvst-5sensors}, bottom) we find earlier beginning of the cooling of the outer sensors $S_1$, $S_2$ and $S_5$ after 40 - 60 s while the inner sensors $S_3$ and $S_4$ are slightly delayed, with significant cooling starting between 60 and 80 s. Further, we observe a series of increasing temperatures from sensor $S_1$ (bottom) to sensor $S_4$ (second from top). However, two peculiarities are observed.

First, the behavior of sensor $S_5$ (top) is slightly unusual as it first cools down at an increased rate, but then levels off. Most likely, this is due to the small distance to the cold container wall. Since at the beginning of the cooling process the sample is in thermal equilibrium there are no temperature or density gradients in the container liquid driving convection. Thus, thermal conduction is the main heat transferring mechanism in the beginning of the cooling process which causes the outer sensors $S_1$ and $S_5$ to cool down first. Then, during cooling, large thermal gradients develop leading to convection which then produce a regular temperature distribution as expected after $\approx$ 200 s.

Second, the cooling of sensor $S_1$ (bottom) is strongly enhanced reaching 0 °C already after $\approx$ 70 s. For our samples we expect ice formation to start at $\approx$ -2 °C\cite{freeze} which then strongly influences the heat transfer. Consistent with the visual observation of the experiment, ice formation took place very quickly at the bottom of the Wolters standard, and which covered sensor $S_1$ after 70 - 75 s. The ice was then further cooled down to a minimum temperature of 210 K after 220 s when the gas flow was cut off to prevent breaking of the Wolters standard.

Thereafter, at $t \approx 240 \textrm{ s}$ the Wolters standard was removed from the cooling chamber resulting in the quick formation of thermal equilibrium within the liquid approximately 1.5 K below the temperature of $S_3$. We continued the measurement and observed convergence of the temperatures of the ice (sensor $S_1$) and liquid (sensors $S_2$ - $S_5$) phases approximately at the expected freezing point of $\approx$ 271 K, as one would expect for liquid-ice mixtures.

Since the cooling curve of sensor $S_3$ which is positioned in the middle of the Wolters standard, is very reproducible (as can be seen from a comparison of the corresponding curves of Fig. \ref{fig:Tvst-5sensors} top and bottom) we take this cooling curve as a measure of performance to compare our newly developed cooling setup to common cooling procedures.

\begin{figure}
\includegraphics[trim=1.5cm 1cm 3cm 2cm,width=0.48\textwidth]{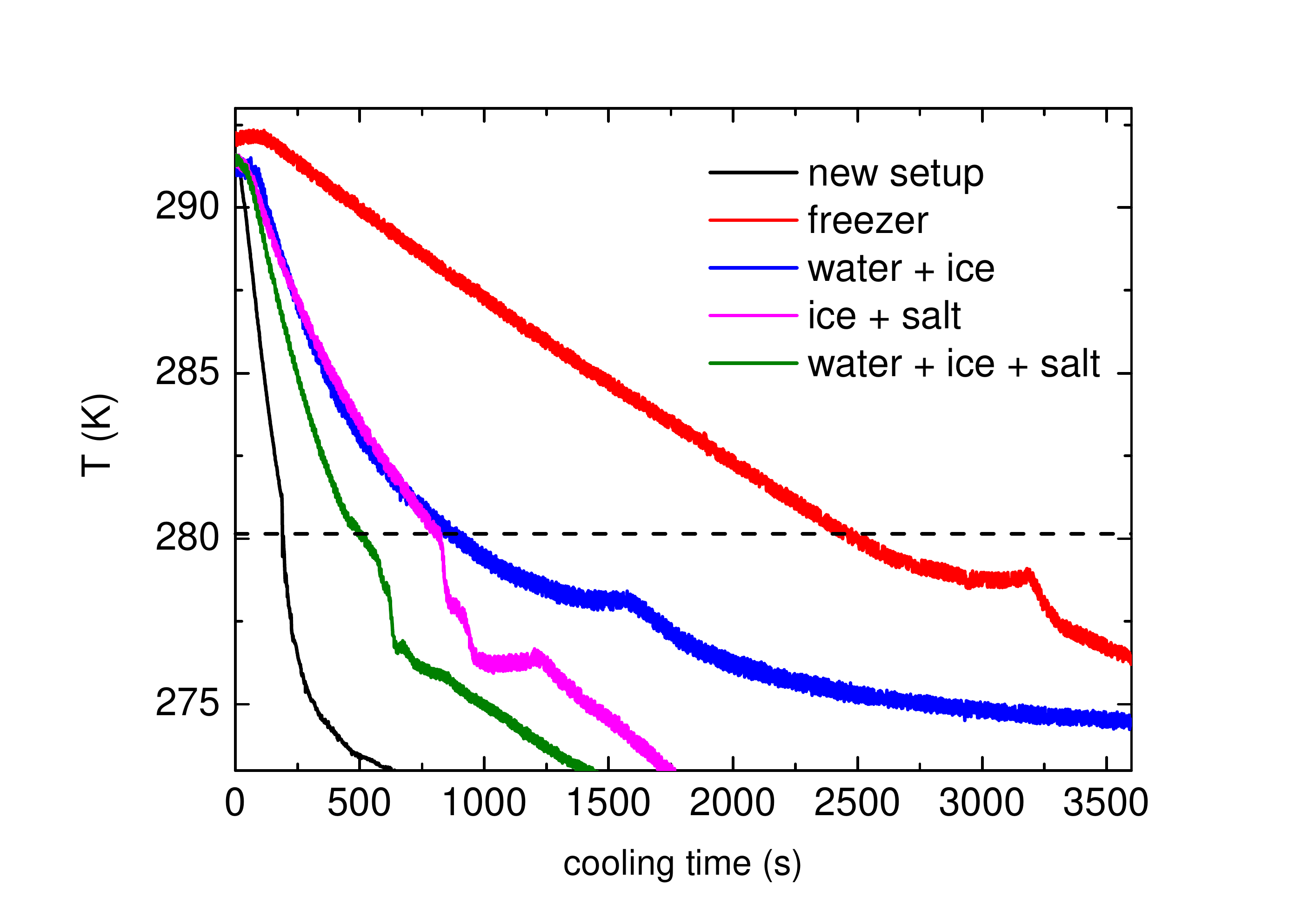}
\caption{(color online) Cooling curves for different common cooling methods compared to our new setup presented here measured with the Wolters standard. All measurements show the time dependence of the $S_3$ thermocouple temperature which is positioned in the middle of the sample container. The dashed black line indicates the desired final temperature of 8 °C and thus determines the cooling time for each method. For further cooling the formation of ice on the sample container walls leads to the anomalies which can be observed for all methods below 280 K.\label{fig:3minutes}}
\end{figure}

We have tested the cooling performance of (a) a household freezer, (b) a water-ice mixture, (c) an ice-salt mixture (about 0,5 l ice cubes and 0,25 kg household salt) and (d) a water-ice-salt mixture with a composition as in case (c) with the empty spaces filled up with water. These methods lead to the following conditions on the outside of our container: (a) cold air at $\approx$ -17°C, (b) water at 0°C, (c) separated ice-glass contact points at $\approx$ -20 °C and (d) full surface contact at $\approx$ -20 °C. A comparison of the cooling curves of temperature sensor $S_3$ in the Wolters standard for these 4 methods is compared to our new cooling setup presented here in Fig. \ref{fig:3minutes}. Since the primary intention was to serve cold beer in a short time we have measured the time to reach the desirable serving temperature of 8°C\cite{8degree1,8degree2,8degree3,8degree4} with each method, and which is indicated by the dashed line in the figure.

Obviously, the simplest method, the household freezer, is the slowest method by far, requiring $\approx$ 2400 s or 40 minutes for the coooling procedure. This result is not surprising since the main part of the heat is transferred via gas which has a low heat capacity and only a temperature of -18 °C. A significant reduction of the cooling time is achieved using either a water-ice mixture ($\approx$ 870 s or 14,5 minutes) or an ice-salt mixture ($\approx$ 810 s or 13,5 minutes). Here the conditions are slightly different since the water-ice mixture has a very high heat capacity and full surface contact which makes the heat transfer from the container surface very efficient, but at a relatively high temperature of 0 °C leading to small temperature gradients. With the salt-ice mixture the temperature outside of the container is significantly reduced down to $\approx$~-20~°C, which strongly increases the temperature gradient by a factor of $\sim$ 2 but at the cost of a reduced surface contact area. Coincidentally, both effects, the reduced temperature and reduced contact area lead to similar cooling times for the water-ice and ice-salt mixtures. Further significant reduction of the cooling time down to $\approx$~500~s or 8,3~minutes can be achieved using a water-ice-salt mixture, since this method combines the advantages of very high heat capacity and a low temperature of -20~°C of the cooling medium with full surface contact.

However, none of the above methods is close to our new cooling setup which is capable of cooling down the Wolters standard to 8°C in $\approx$ 190 s or 3,2 minutes. Thus, we have achieved a massive reduction of the cooling time down to between 8\% (freezer) and 38\% (water-ice-salt-mixture) of the cooling time of other common cooling methods. This result indicates that the effect of the very large temperature gradients involved in our setup outweighs the effect of the large heat capacity of a dense coolant such as water by far. Since we are close to the breaking limit of the container material due to thermally induced mechanical stress, neither further increase of temperature gradients nor an increased heat capacity of the coolant is desirable. Thus, we conclude that in terms of minimization of the cooling time with the given sample containers, our setup is already very close to the optimum cooling procedure as it is possible with reasonable efforts.

\section{Simulation}

In order to theoretically support our experimental observations on temperature change and distribution of consumable liquids in our cooler, we have simulated the heat flow of our setup using \emph{Comsol 5.2} with laminar flow and heat transfer in liquids packages. To keep the calculation efforts within reasonable limits, we have made some simplifying assumptions. First, we have modelled only one liquid container in two dimensions as a disc with outer diameter of 6.76 cm and a wall thickness of 3.55 mm\footnote{The Wolters standard glass thickness and diameter varies notably. We have measured a diameter of 67.64(34)~mm and a thickness of 3.55(42)~mm as a result of 9 and 5 measurement points, respectively.}. Thus, we simulate an infinitivly long cylinder which is a good representation of our bottle body when neglecting neck and bottom, hence, for the most part of the liquid. Since the distance between neighboring bottles in the cooling chamber varies strongly from neck to bottom we have estimated an average distance of 4 cm. Thus, the area of free gas flow besides the simulated bottle was set to half of this value since such a geometry resembles real conditions.

For the simulations, the boundary conditions were simplified in the following way: For the filled cooling chamber we observe a temperature profile as a function of height above the metal support grid which ranges from 80~K at the bottom to 110~K at the top of the bottle, i.e., 6~cm above (compare Fig. \ref{fig:TvsPos-cooler}). In our simulations, we have set the temperature of the incoming gas flow to the average value of 95 K. Since our liquid is at room temperature the small differences in the temperature gradients due to this simplification are negligible. 

From a 20 min cooling experiment with the gas flow adjusted at a level as low as possible with the installed manual valve, we have estimated a lower limit for the  liquid nitrogen consumption of about 30 ml s$^{-1}$. For our simulation, and in our geometry this results in an upward gas flow of $\geq$~2,3 cm s$^{-1}$. Since usually the experimental adjustment is not performed at precisely this lower limit, careful increased gas flow rates in a range somewhat above are reasonable. Therefore, in the simulation we have set this value to 10 cm s$^{-1}$, and which in the end results in good agreement between simulation and experiment.

Heat capacity and thermal conductivity of our sample consisting of 95\% water and 5\% ethanol were  treated as temperature independent, since temperature changes within the liquid are well below 20 K. For glass and nitrogen these properties were modelled as functions of temperature due to the high thermal gradients expected within these materials. All values are defined as published in Refs. \cite{lb,lb2}.

We have started by carrying out first simulation runs only taking into account heat conduction. However, these simulations resulted in far too slow cooling rates. In particular, such simulations result in temperature changes well below 1 K at the middle sensor 3 after 3 minutes. To achieve a better agreement between theory and experiment assuming heat conduction appears only possible assuming unreasonably large thermal conductivities or gas flow rates. Hence, we can rule out a model limited to heat conduction. Instead, as already suggested by the temperature-time-dependencies depicted in Fig. \ref{fig:Tvst-5sensors}, we have to take into account convection within our liquid.

In consequence, we have included convection as a gravitational body force in our simulation. For simplicity, and as first approximation, we assume laminar flow for the convective liquid. Then, to obtain good agreement between our spatially and temporally resolved experimental data and the numerical simulation we have to increase the body forces driving the convection by a factor of $\approx$~4. A physical interpretation of this required increase will be given in the discussion section.

\begin{figure}
\includegraphics[trim=1.5cm 1cm 3cm 2cm, width=0.48\textwidth]{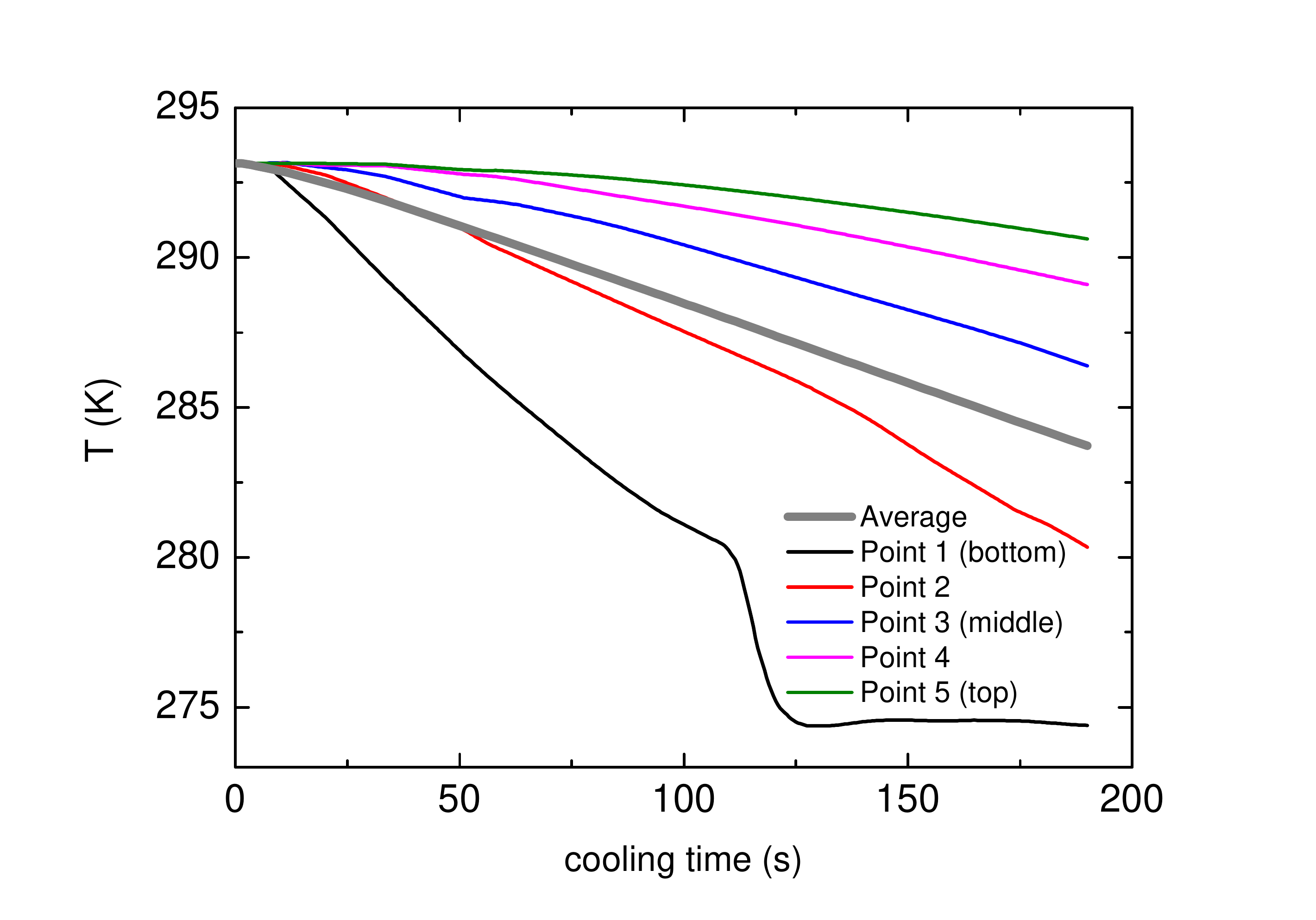}\raisebox{0.06\textwidth}{\makebox[0pt][r]{\includegraphics[width=0.15\textwidth]{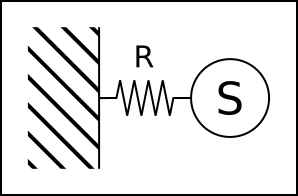}\hspace{0.24\textwidth}}}
\caption{(color online) Temperature vs. cooling time as calculated from our simulation. The points 1 - 5 are selected as the thermocouple positions in the experiments with vertical alignment. Colors correspond to the respective sensor colors as in Fig. \ref{fig:Tvst-5sensors}, bottom. In addition, the thick gray line represents the average temperature of the liquid. The inset depicts the thermal basic model used for the simulations. The hatched area is the temperature bath at 95 K, $R$ denotes the thermal resistivity of the container material and $S$ is the liquid sample; for details see text.\label{fig:sim}}
\end{figure}

As a result from our simulation we find the time and spatially resolved temperature and flow distributions of our whole setup. In Fig. \ref{fig:sim} we have plotted the temperature evolution for 5 selected points ($P_1$ to $P_5$) as well as the average temperature of the liquid. The points $P_1$ to $P_5$ were taken at the positions of the thermocouples in our experiments with vertical alignment of the Wolters standard. Hence, in case of perfect agreement between experimental and numerical data Figs. \ref{fig:Tvst-5sensors}, bottom and \ref{fig:sim} should be identical. We recall that in Fig. 4 the temperature evolution of Sensor $S_1$ was affected by ice formation. Thus, this sensor should not be considered in the comparison to the simulation.

Then, from a comparison we recognize that both figures agree nicely for the sensors $S_2$ to $S_5$. However, some differences may be noted: First, in the simulation a decreasing temperature at $P_1$ is observed after $\approx$~10~s, while the corresponding experimental temperature of $S_1$ remains constant for the first 30~s. Similarly, for the points $P_2$ - $P_5$ significant cooling can be observed $\approx$~20~s prior to the corresponding thermocouples. Still, this discrepancy is relatively small compared to the total experimental time of $\approx$~200~s. Possible explanations for the difference might be inaccurate timing of our experimental data, slightly different material parameters of our Wolters standard compared to the literature values used, misalignment of points $P_1$ to $P_5$ with respect to the corresponding sensors or a combination of all of the above.

Second, the simulated temperatures of all five points form a continuously rising series from bottom to top at all times. In contrast, the experimentally observed temperature of sensor 5 remains slightly below the temperature of sensor 4 up to a time of $\approx$ 200~s. We attribute this difference to small amounts of air trapped inside the Wolters standard which alters the thermodynamic properties at the top (close to sensor 5) and which is neglected in the simulation.

Altogether, while there are minor differences between experimental and numerical data, three essential features are reproduced very well: (i) First, and most notably, we find an absolute cooling of our liquid of $\approx$~10~K in 180~s. (ii) Consistently, the average temperature of the liquid lies about $\approx$~1.5~-~2~K below the temperatures of $S_3$ / $P_3$. And (iii), the time-evolution of the $S_2$ - $S_5$ and $P_2$ - $P_5$ temperatures depicted in Figs. \ref{fig:Tvst-5sensors} (bottom) and \ref{fig:sim} is very similar. Altogether, we conclude that our simulation of our cooling process describes the essential behavior our setup very well.


\section{Results and discussion}

As we have discussed above, the optimization of the cooling setup in terms of modifying the cooling chamber appears hardly possible. From the good agreement between our simulated and experimental data we conclude that the thermal physics of our setup is essentially determined by heat conduction and convection. These flows are either forced as our steady flow of cold convective nitrogen gas or driven by forces within the container liquid, that is by gravity, acting upon the liquid with its different local densities and leading to convection.

\begin{figure}
\includegraphics[width=0.48\textwidth]{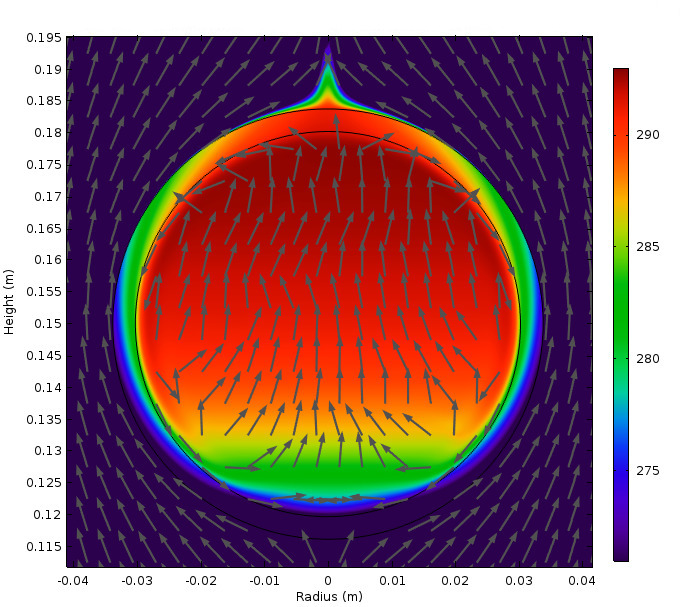} \includegraphics[width=0.48\textwidth]{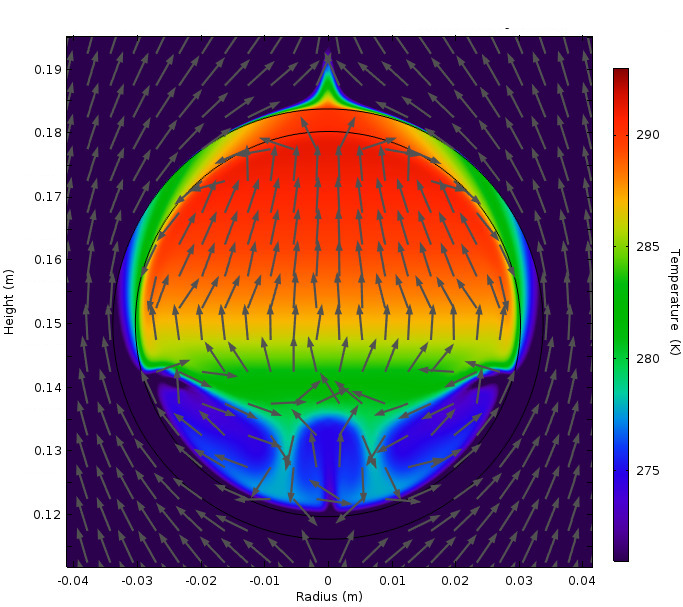}
\caption{(color online) Temperature and material flow distributions in the Wolters standard after cooling times of 80 (top) and 180~s (bottom) as resulting from the numerical simulation. The temperature color range is chosen from 271~K (estimated freezing point) to 293~K (starting temperature of the liquid). Parts of the simulation, in particular the cold nitrogen gas, are out of this range. Arrows indicate the local flow directions and reveal the formation of multiple vortices in the liquid. Maximum velocities in the liquid phase during the cooling process are 10 mm s$^{-1}$.
\label{fig:sim-cross}}
\end{figure}

When comparing our simulation to the experiment, there are in particular two peculiarities of our setup which we want to discuss in further detail: First, from the temporal evolution of our experimental temperature data depicted in Fig. \ref{fig:Tvst-5sensors} we have concluded that we need to take convection into account, since the heat flow within the liquid is strongly increased compared to conductive heat flow alone. However, a priori, a simple assumption of convection should lead to an overall increased material flow but similarly in all directions. Contrary to this expectation, we observe almost identical temperatures in horizontal alignment (Fig. \ref{fig:Tvst-5sensors}, top) while a significant temperature gradient is maintained at all times for the vertical alignment (Fig. \ref{fig:Tvst-5sensors}, bottom). From our simulation data we find a possible explanation for this observed anisotropy. In Fig. \ref{fig:sim-cross} we have depicted the temperature distributions and flow directions of our simulated data after cooling times of 80 and 180~s, qualitatively representing the thermal and flow conditions in the beginning and at the end of the cooling process.

In the beginning of the simulation up to $\approx$~100~s we observe the conditions depicted in Fig. \ref{fig:sim-cross} (top) with a fast formation of a growing surface layer on the inside container wall with a thickness of only few millimeters. In this layer a strong flow of up to $\approx$~10~mm~s$^{-1}$ downward along the container wall is observed, and which results in strong cooling at the container bottom while no convective heat transport in vertical direction occurs. At the bottom of the container the cold flow from the container walls mixes with the warm liquid further inside and a homogenuous upward flow across the whole central part of the liquid develops. This leads to the very homogenuous vertical temperature distribution with a significant temperature gradient developing in vertical direction. Obviously, our temperature sensors $S_1$ and $S_5$ are inside the homogenuous upward flow area since we do not observe signs of the cold downward flowing surface layer.

After $\approx$~100~s this simple flow pattern evolves as depicted in Fig. \ref{fig:sim-cross} (bottom). At this point a significant fraction of the liquid reaches a temperature of $\approx$~4~°C where the density has a maximum due to the high water content. This cold liquid fraction forms small new vortices at the bottom of the container. The flow conditions in the warmer upper part remain as before with the border between vortices and the homogenuous zone moving upward upon further cooling. This new arrangement of vortices leads to the formation of growing areas of liquid at $\approx$~4~°C at the bottom and thus to anomalies as observed in Figs. \ref{fig:3minutes} and \ref{fig:sim} for sensor $S_3$ and point $P_5$ at temperatures below 280~K, respectively. Hence, the simulation appears to produce explanations for all time dependent temperature measurements, including the observed anomalies.

Conceptually, for our sample with a high water content the density anomaly at 4~°C is very beneficial since it leads to the development of multiple vortices at the bottom of the container and thus to significant convection also at the coldest spots within the liquid. In contrast, for liquids without such density anomaly, we expect ice formation at the bottom to start faster, reducing the cooling performance and increasing the final temperature achievable without modifications of the cooling setup. However, a strong reduction of cooling times compared to common methods can be expected for water-free liquids as well.

The second remarkable peculiarity is our simulations body force (gravity), which needs a factor of 4, to achieve good agreement with the experimental data. Our interpretation of this factor is that we have a strongly increased liquid flow due to additional forced convection in our setup. This increase is most likely a result of the vibrations in the cooling chamber which are caused by the turbulent flow of cold nitrogen gas and droplets from the silcone hose. These vibrations then lead to enhancement of the material flow on very small spatial scales, which one might think of as ``artificial turbulence`` in the liquid. Since the vibrations have frequencies of some hertz at fluid flow rates of some millimeters per second, a very rough estimate of a spatial scale for such artificial turbulence would be about 1~mm and below. We rule out modifications of the materials flow on larger scales since (except for the factor of 4 in the body forces) the laminar flow assumend and calculated in our simulation is in very good agreement with all of our experimental data.

In conclusion, we have presented a very simple setup for cooling of liquids in sealed containers and at intermediate temperatures, that is close to their freezing point and above. In particular, our setup does not require electrical power which possibly might be a requirement for specific applications. Due to the simplicity of our setup all components can be easily adapted for different types of containers, liquids and temperature ranges as well as scaled up or down. We have measured time dependent temperature distributions of the main components and compared these to numerical data from FEM modelling for which we find very good agreement. Thus, the physics of our system can be understood taking into account heat and material flow in the container and container liquid.

From the analysis of our data we find that our cooling chamber design is quite close to the optimum that can be reached with reasonable efforts. For heat transport within the container and liquid we find a limitation of heat flow in the beginning of the cooling process, that is in the first $\approx$~20~s, due to the finite thermal conductivity of the container material. Later, fast heat transport within the liquid is crucial which is dominated by convection caused by the large temperature gradients involved. However, from a comparison of simulation and experiment we find a significant increase of convection which we attribute to strong vibrations of the cooling chamber and thereby induced turbulence.

The latter aspects make room for further improvements of the cooling performance of our setup by choosing different container materials (which we did not follow up on since our intention was to cool the Wolters standard) or by further increasing convection. Such increase of convection in the liquid might be possible by increasing vibrations or somehow inducing random movements. Another approach could be to generate ``controlled convection'' as for example with the SpinChill system where a sealed containers is rotated in a temperature bath electrically\cite{spinchill}. However, the trade-off of such improvements is most likely the need for additional electrical components or more sophisticated mechanical design. Since for our application the achieved cooling time of $\approx$~3 minutes is has been considered satisfactory by the experimental team, we have not further tested such improvements.




%
%

%

\begin{acknowledgments}
We would like to acknowledge the special support from Andreas Hördt from the Institute for Geophysics and Extraterrestial Physics at the TU Braunschweig, who permitted us access to the institutes Comsol simulation environment to perform the calculations presented here.

Furthermore, we want to emphasize important contributions from former lab members. First, we want to thank Matthias Bleckmann who came up with the idea of a regular informal after-work-meeting and thus induced the need for regular fast cooling. He also took part in the development of the first working prototypes. And second, we want to thank Ali Awada who often joined these meetings and helped in the construction of the final cooling setup presented here.
\end{acknowledgments}


\begin{thebibliography}{11}
\bibitem{theunissen94} M. A. Theunissen (borrelmeister), private comm. (1994).
\bibitem{amitsuka13} H. Amitsuka, private comm. (2013).
\bibitem{sun01} D.-W. Sun, \textit{Advances in food refrigeration}, Leatherhead: Leatherhead Food RA Pub. (2001). 
\bibitem{chapman30} J. Chapman Hilder, Popular Science Monthly, September 1930, pp. 26 - 27. 
\bibitem{kittel} C. Kittel, \textit{Introduction to Solid State Physics 1}, 8th Ed., John Wiley \& Sons, Inc. (2005).
\bibitem{spinchill} SpinChill Website, \verb|http://www.spinchill.com|, 05/26/2016.
\bibitem{lb} W. Dienemann \textit{et al.} in Landolt-Börnstein - 4. Teil Bandteil a, Springer-Verlag (1967)
\bibitem{lb2} F. Bo$\breve{\textrm{s}}$njakovi$\acute{\textrm{c}}$ \textit{et al.} in Landolt-Börnstein - 4. Teil Bandteil b, Springer-Verlag (1972)
\bibitem{freeze} F. M. Raoult, Comptes rendus \textbf{90}, p. 865 (1880). 
\bibitem{8degree1} Standort38.de Website, \emph{www.standort38.de/de/unter nehmen/freizeit-gesundheit/Bier-ist-das-ideale-Grill-Getraenk.aspx}, 11/11/2016
\bibitem{8degree2} Deutscher Brauer Bund e.V. Website, \emph{http://www.brauer-bund.de/bier-ist-genuss/optimale-trinktemperatur.html}, 11/11/2016
\bibitem{8degree3} Bier ABC Website, \emph{http://www.das-bier.net/biertemperatur.html}, 11/11/2016
\bibitem{8degree4} Bier Entdecken Website, \emph{http://www.bier-entdecken.de/die-richtige-trinktemperatur/}, 11/11/2016
\bibitem{beer-pressure} F. Steinki (workshop employee, local brewery), private comm. (2016)
\end{thebibliography}

\end{document}